\documentclass[mathpazo]{cicp}

\usepackage{amsmath}
\usepackage{enumerate}
\usepackage{multirow}
\usepackage{graphicx,subfigure}

\begin{document}
\title{Numerical Methods and Comparisons for 1D and Quasi 2D Fluid Streamer Propagation Models}



 \author[Mengmin Huang et.~al.]{Mengmin Huang$^1$, {Chijie Zhuang}\corrauth$^2$, Huizhe Guan$^2$, Rong Zeng$^2$}
 \address{$^1$Department of Mathematics, National University of Singapore, Singapore 119076.\\$^2$Department of Electrical Engineering, Tsinghua University, Beijing 100084, China.}
 \emails{{\tt Mengmin.Huang@cgg.com} (Huang), {\tt chijie@tsinghua.edu.cn} (Zhuang), {\tt ghz2009@live.cn} (Guan), {\tt zengrong@tsinghua.edu.cn} (Zeng)}

\begin{abstract}
In this work, we propose and compare four different strategies to simulate the fluid model for streamer propagation in one-dimension (1D) and quasi two-dimension (2D), which consists of a Poisson's equation for particle velocity and two continuity equations for particle transport. Each strategy involves of one method for solving Poisson's equation and the other for solving continuity equations, and a total variation diminishing three-stage Runge-Kutta method in temporal discretization. The numerical methods for Poisson's equation include finite volume method, discontinuous Galerkin methods, mixed finite element method and least-squared finite element method. The numerical method for continuity equations is chosen from the family of discontinuous Galerkin methods. The accuracy tests and comparisons show that all of these four strategies are suitable and competitive in streamer simulations from the aspects of accuracy and efficiency. Results show these methods are compatible. By applying any strategy in real simulations, we can study the dynamics of streamer propagations in both 1D and quasi 2D models.
\end{abstract}


\maketitle

\section{Introduction}
Streamers is a type of electrical discharge emerging when a strong electric field is applied to a gap, e.g., an air gas. It occurs in nature as well as in many industrial applications such as ozone generation, air purification and plasma assisted combustion. Since the streamers develops within a short time, e.g., nanoseconds or microseconds, it is difficulty to measure all the micro physical parameters by experiment, which leads us to use numerical simulation to study the physics of streamer during the last several decades\cite{dh1,wu,be,du,ge,min1,mon,pan,zhuang2, zhuang1,zhuang}.

The simplest model for simulating the streamer propagation is the fluid model, which involves of two continuity equations for particle densities coupled with Poisson's equation for electric potential and electric field.
\begin{equation}\label{1dmodel}
  \begin{cases}
      \partial_t\sigma+\partial_z(\mu_{\sigma}\sigma E)-D\partial_z^2\sigma=S|E|e^{K/|E|}\sigma, & \text{ in }(0,1)\times(0,T); \\
      \partial_t\rho+\partial_z(\mu_{\rho}\rho E)=S|E|e^{K/|E|}\sigma, & \text{ in }(0,1)\times(0,T); \\
      -\partial_z^2\phi=\rho-\sigma, E=-\frac{\partial\phi}{\partial z}, & \text{ in }(0,1)\times(0,T).
    \end{cases}
\end{equation}
In this model, $\sigma$, $\rho$, $\phi$ and $E$ are rescaled electron density, positive ion density, electric potential and electric field respectively. $\mu_{\sigma}=-1$, $\mu_{\rho}$ are rescaled mobility constant for electron and positive ion, respectively. $D$ is the rescaled diffusion coefficient for electron. The rest two rescaled parameters, $S$ and $K$, are defined as $S=APx_0$ abd $K=\frac{BPx_0}{V_0}$,
where $x_0$ and $V_0$ are length scale and applied potential scale respectively, $P$ is the pressure in torr and $A$, $B$ are two constants.

Dirichlet boundary conditions are imposed for Poisson's equation, i.e., $\phi(0,t)=0$ and $\phi(1,t)=1\text{ or }-1$. Homogeneous Neumann boundary conditions are imposed for continuity equations as we allow the fluxes of particles to pass through the boundaries. The initial data is often assumed to be a Gaussian function which describes the particle distribution after the electron avalanche.

The so-called quasi 2D model is as follows,
\begin{equation}\label{q2dmodel}
  \begin{cases}
      \partial_t\sigma+\frac{1}{r}\partial_r(r\mu_{\sigma}\sigma E)-\frac{D}{r}\partial_r(r\partial_r\sigma)=S|E|e^{K/|E|}\sigma, \\
      \partial_t\rho+\frac{1}{r}\partial_r(r\mu_{\rho}\rho E)=S|E|e^{K/|E|}\sigma, \\
      -\frac{1}{r}\frac{\partial}{\partial r}\left(r\frac{\partial\phi}{\partial r}\right)=\rho-\sigma, E=-\frac{\partial\phi}{\partial r}.
    \end{cases}
\end{equation}
The notations in quasi 2D model are same as those in 1D model. Quasi 2D model can be derived from a 2D model with central symmetry by using polar coordinates to change the spatial variables. The computational domain can generally be assumed as $\Omega=[r^0,1]$. The boundary conditions for continuity equations are still the homogeneous Neumann type. But the boundary conditions for Poisson's equation are set up distinctively in two different cases.

\textbf{Case 1}, $r^0=0$. In this case, the computational domain before changing of variable is a disc which includes the origin $r=0$. Thus, the boundary conditions for Poisson's equation are given by the following. At $r=0$, we impose Neumann boundary condition $\frac{\partial\phi}{\partial r}=0$ to avoid irregularity; and at $r=1$, a Dirichlet boundary condition is imposed to assure the well-poseness. In fact, there is no truly physical application in such case. In this work, this case is used to test and compare our algorithms and to study the extensions of our method to quasi three dimensional model \cite{zhuang2}.

\textbf{Case 2}, $r^0>0$. In this case, the domain is a ring which excludes the origin. It is possible to impose Dirichlet boundary conditions, e.g., $\phi(r^0,t)=0$ and $\phi(1,t)=1\text{ or }-1$.

The continuity equations in the above two models are convection dominated if the source terms are not taken into consideration. As is well known, the traditional linear finite difference schemes for convection equations usually generate too many numerical oscillations or diffusions \cite{mor}. Hence, a numerical method free of numerical oscillation and diffusion is desired. In addition, it has been found that the solution of streamer model has steep derivatives or even has discontinuities under some configurations. Therefore, a good numerical method should be of high resolution and be able to capture the sharp changes.

Many numerical methods have been proposed to solve the streamer models. Flux corrected transport (FCT) technique \cite{bor,boo,za} was applied to finite difference method (FDM) to overcome the drawback of traditional linear finite difference scheme during 1980s and 1990s \cite{dh1,mor,wu}. However, it is hard for FDM to handle the unstructured meshes or complex geometries. Therefore, starting from 1990s, FCT was been combined to finite element method (FEM) \cite{ge1,min1}. The good news for FEM-FCT was that linear and nodal-based FEM can maintain a comparable accuracy as FDM-FCT and was easy to implement. But on the other hand, FEM cannot guarantee the local conservation \cite{zhuang}; thus, the total current law on electromagnetism is violated. To enforce local conservation, finite volume method (FVM) becomes popular since 2000 \cite{be,du,mon,pan}. Although FVM can also handle complex geometries, it needs wide stencil to construct high order scheme which can make computation inefficient.

From the above literature review, our purpose is to find out some numerical methods which is of high resolution, is able to avoid non-physical solution, can preserve the local conservation and can be easily extended to complex geometries and unstructured meshes. With the help of such methods, we may simulate the streamer propagation process accurately and capture the physical properties.

To achieve this goal, we apply the so-called Oden-Babu$\check{s}$ka-Baumann discontinuous Galerkin (OBBDG) method \cite{daw,obb,ri} and local discontinuous Galerkin (LDG) method \cite{co,co1,co2,zhuang1}. Both of them are from the class of discontinuous Galerkin (DG) methods which use finite element space discretization but allows the solution to have discontinuities along the interface of adjoint elements. Consequently, these two methods can enforce the local conservation, achieve high accuracy and handle the complex regions; in the other words, they possesses the advantages of FEM and FVM. Besides, these methods can control the numerical oscillations with the help of a slope limiter. In this paper, we will show that LDG and OBBDG methods are both good competitors in simulations of streamer propagation.

So far, we have discussed the numerical methods for continuities equation. For the Poisson's equation, there are many existing methods which can solve it very well. For instance, the finite volume method used by U. Ebert, D. Bessi$\grave{e}$res \emph{et al}. \cite{be,mon} and discontinuous Galerkin method introduced by D. Arnold, M. Wheeler \emph{et al}. \cite{ar,whe}. Both of these two methods directly solve the Poisson's equation, and use the derivative of the numerical solution to approximate the rescaled electric field in continuity equations. Since it is the electric field coupled with continuity equations rather than electric potential, it is a natural idea to seek some numerical methods which can directly derive a solution of high accuracy for electric field. To achieve this goal, we refer to the mixed finite element method (MFEM) \cite{br,br1} and least-squares finite element method (LSFEM) \cite{bo,bo1}. Both methods rewrite the Poisson's equation as a first order equation system where the electric potential and field becomes two independent variables, called scalar and flux variable respectively. The difference is that the choices for the finite dimensional subspaces for both variables in MFEM should satisfy the inf-sup condition while the choices are independent in LSFEM.

In summary, there are two main purpose in this work. One is to combine one of the numerical methods for Poisson's equation with LDG or OBBDG method to form a strategy to solve the governing system \eqref{1dmodel} or \eqref{q2dmodel} and then to test and compare the performances of different strategies. The other purpose is to apply the best strategy to some simple simulations of negative streamer propagation to study the dynamics and influences due to the change of parameters in the systems.

This paper will be organized as follows. The numerical methods in 1D model and quasi 2D model will be discussed in Section 2 and Section 3 respectively. In both section, we will firstly introduce the different numerical schemes for continuity equations and Poisson's equation; and then, numerical comparisons among some combinations of methods will be given; finally, some simulation results will be shown by picking one of the combinations.

\section{Numerical Methods in 1D model}

Let us consider Model \eqref{1dmodel}. Suppose the time step size is $\tau$, the numerical algorithm is designed as follows: assume at any time level $t^n=n\tau$, we have the numerical solutions for particle densities, $\sigma^n$ and $\rho^n$, then we use $\sigma^n$ and $\rho^n$ to solve the Poisson's equation numerically to obtain $\phi^n$; after that, we plug a proper numerical approximation of $E^n$ into continuity equations to solve for $\sigma^{n+1}$ and $\rho^{n+1}$. This process will repeat until the simulation finishes.

Let $0=z_0<z_1<\cdots<z_N=1$ be a uniform spatial partition of computational domain $[0,1]$ such that $z_j=jh$ where $h=\frac{1}{N}$ for $j=0,1,\cdots,N$. Denote the subintervals by $I_j=[z_j,z_{j+1}]$, $j=0,1,\cdots,N-1$. Let $N_i$, $N_d$ and $N_n$ denote the sets of labels of interior, Dirichlet boundary and Neumann boundary nodes respectively.

\subsection{Discontinuous Galerkin Method for Continuity Equations}

As mentioned above, we apply Oden-Babu$\check{s}$ka-Baumann discontinuous Galerkin (OBBDG) method and local discontinuous Galerkin (LDG) method from the DG class to solve the continuity equations.

Denote the finite dimensional space by
$$V_k=\{v : v|_{I_j}\in \mathbb{P}_k(I_j),\text{ for }j=0,1,\cdots,N-1\},$$
where $\mathbb{P}_k(I_j)$ is the space of polynomials of degree up to $k$ on $I_j$.

The numerical solutions are allow to have discontinuities at the interior nodes, and we define the average $\{v\}$, and jump $[v]$, of $v$ at each interior node $z_j$,
\begin{equation}
  \{v\}=\frac{1}{2}[v(z_j^-)+v(z_j^+)],\;[v(z_j)]=v(z_j^-)-v(z_j^+),\;\forall j=1,2,\cdots,N-1,
\end{equation}
where $v(z^{\pm})=\displaystyle\lim_{\epsilon\to 0^+}v(z\pm\epsilon)$. We extend the definition of average and jump to the endpoints as well,
\begin{equation}
  [v(z_0)]=-v(z_0^+),\;\{v(z_0)\}=v(z_0^+),\;[v(z_N)]=v(z_N^-),\;\{v(z_N)\}=v(z_N^-).
\end{equation}

\subsubsection{The OBBDG method}

The OBBDG method is to find $\sigma^h(z,t)$ and $\rho^h(z,t)\in V_k$ such that for $t=0$,
\begin{equation}
  \int_0^1(\sigma^h(z,0)-\sigma(z,0))v=\int_0^1(\rho^h(z,0)-\rho(z,0))v=0,\;\forall v\in V_k;
\end{equation}
and for $t=t^n>0$,
\begin{equation}\label{obb}
  \begin{aligned}
    &\int_0^1\partial_t\sigma^hv+C(\sigma^h,v;E^n)+B(\sigma^h,v)=L(\sigma^h,v;E^n),\;\forall v\in V_k, \\
    &\int_0^1\partial_t\rho^hv+C(\rho^h,v;E^n)=L(\sigma^h,v;E^n),\;\forall v\in V_k.
  \end{aligned}
\end{equation}
where $C(\sigma^h,v;E^n)$, $C(\rho^h,v;E^n)$, $B(\sigma^h,v$ and $L(\sigma^h,v;E^n)$ are explained below.
In \eqref{obb}, $C(\sigma^h,v;E^n)$ and $C(\rho^h,v;E^n)$ are discretization scheme for the convection terms,
\begin{eqnarray*}
  C(P^h,v;E^n)&=&-\sum_{j=0}^{N-1}\int_{I_j}P^h\mu_PE^n\frac{dv}{dz} \\
  &&+\sum_{j\in N_i}\widehat{P^h}(z_j)\mu_PE^n(z_j)[v(z_j)]+\sum_{j\in N_n}[P^h(z_j)\mu_PE^n(z_j)v(z_j)],
\end{eqnarray*}
where we apply an upwind-type numerical flux
$$\widehat{P^h}(z)=\begin{cases}
  P^h(z^-) & \text{ if }\mu_PE^n(z)\geq 0 \\
  P^h(z^+) & \text{ if }\mu_PE^n(z)< 0
  \end{cases},$$
for $P=\sigma$ or $\rho$. $B(\sigma^h,v)$ is the discretization scheme for the diffusion term of $\sigma$,
$$B(\sigma^h,v)=\sum_{j=0}^{N-1}\int_{I_j}D\frac{d\sigma^h}{dz}\frac{dv}{dz}-\sum_{j\in N_i}\left\{D\frac{d\sigma^h}{dz}(z_j)\right\}[v(z_j)]+\sum_{j\in N_i}\left\{D\frac{dv}{dz}(z_j)\right\}[\sigma^h(z_j)].$$
Finally, $L(\sigma^h,v;E^n)$ is the discretization scheme for the source term,
$$L(\sigma^h,v;E^n)=\int_0^1S|E^n|e^{K/|E^n|}\sigma^hv.$$

\subsubsection{The LDG method}

The diffusion term is directly discretized in the OBBDG method. In the LDG method, an auxiliary variable is introduced to convert diffusion term to convection term; and the new equation for the auxiliary variable is also of first order.

More precisely, the auxiliary variable is
$$q=\frac{\partial\sigma}{\partial z},$$
then the LDG method is to find $\sigma^h(z,t),\rho^h(z,t),q^h\in V_k$ such that for $t=0$,
\begin{equation}
  \int_0^1(\sigma^h(z,0)-\sigma(z,0))v=\int_0^1(\rho^h(z,0)-\rho(z,0))v=0,\;\forall v\in V_k;
\end{equation}
and for $t=t^n>0$, for each element $I_j$,
\begin{equation}\label{ldg1}
  \int_0^1q^hv=\sum_{j=0}^N\widehat{\sigma^h}(z_j)[v(z_j)]-\int_0^1\sigma^h\frac{dv}{dz},\;\forall v\in V_k,
\end{equation}
\begin{eqnarray}\label{ldg2}
\int_0^1\partial_t\sigma^hv&+&\sum_{j=0}^N(\mu_{\sigma}E^n(z_j)\widetilde{\sigma^h}(z_j)-D\widehat{q^h}(z_j))[v(z_j)]\nonumber \\
&&-\int_0^1(\sigma^h\mu_{\sigma}E^n-Dq^h)\frac{dv}{dz}=\int_0^1S|E^n|e^{K/|E^n|}\sigma^hv,\;\forall v\in V_k,
\end{eqnarray}
\begin{eqnarray}\label{ldg3}
  \int_0^1\partial_t\rho^hv&+&\sum_{j=0}^N\mu_{\rho}E^n(z_j)\widetilde{\rho^h}(z_j)[v(z_j)]\nonumber \\
  &&-\int_0^1\rho^h\mu_{\rho}E^n\frac{dv}{dz}=\int_0^1S|E^n|e^{K/|E^n|}\sigma^hv,\;\forall v\in V_k.
\end{eqnarray}
In \eqref{ldg1}-\eqref{ldg3}, the numerical flux in convection terms is defined by upwind type,
$$\widetilde{P^h}(z)=\begin{cases}
  P^h(z^-) & \text{ if }\mu_PE^n(z)\geq 0 \\
  P^h(z^+) & \text{ if }\mu_PE^n(z)< 0
  \end{cases},$$
for $P=\sigma$ or $\rho$. The numerical fluxes, $\widehat{\sigma^h}$ and $\widehat{q^h}$, defined in discretization of diffusion term and auxiliary equation are chosen according to the alternating principle, i.e.,
$$\widehat{\sigma^h}(z)=\sigma^h(z^+),\;\widehat{q^h}(z)=q^h(z^-),$$
or
$$\widehat{\sigma^h}(z)=\sigma^h(z^-),\;\widehat{q^h}(z)=q^h(z^+).$$

\subsubsection{The slope limiter}

As mentioned above, a slope limiter is desired to collaborate with the discontinuous Galerkin schemes to avoid nonphysical solutions. A slope limiter proposed by Krivodonova \cite{kri} will be applied in our work. To illustrate this slope limiter, we firstly assume the numerical solution in the element $I_j$ can be presented by
\begin{equation}\label{jspan}
  U_j=\sum_{l=0}^{p}c_{j,l}P_l(\xi),
\end{equation}
where $P_l$ is the $l$-th order Legendre polynomial and $\xi=\frac{z-(j+1/2)h}{h/2}$.

The slope limiter works from the highest order coefficient in \eqref{jspan} to the lowest order coefficient. It replaces $c_l$ with
$$\hat{c}_{j,l}=\mbox{minmod}(c_{j,l},\alpha_l(c_{j+1,l-1}-c_{j,l-1}),\alpha_l(c_{j,l-1}-c_{j-1,l-1})),$$
where the parameter $\alpha_l$ satisfies
$$\frac{1}{2(2l-1)}\leq \alpha_l\leq 1,$$
and the minmod function is defined by
$$\mbox{minmod}(a,b,c)=\begin{cases}
  s\min\{|a|,|b|,|c|\} & \text{if }s=\mbox{sign}(a)=\mbox{sign}(b)=\mbox{sign}(c), \\
0 & \text{otherwise}.
\end{cases}$$
In practice, the parameter $\alpha_l$ is set to be 1 to make the numerical solution least diffusive. The slope limiter will not stop until $\hat{c}_{j,l}=c_{j,l}$ for some $l$ or $l=1$ \cite{kri}. Note that the lowest order coefficient does not need to be limited because of the orthogonality of Legendre polynomials.

\subsubsection{Fully discrete formulation}

To seek for a good spatial discretization for continuity equations, we take care about the dominant convection term. In addition, we apply a third order total variation diminishing (TVD) Runge-Kutta method (TVDRK3) \cite{shu} in temporal discretization.

In the LDG method, the auxiliary variable $q^h$ can be solved by using $\sigma^n$ \eqref{ldg1} from element to element and then we can plug $q^h$ into \eqref{ldg2} to solve $\sigma^{n+1}$. Thus, in summary, taking $\sigma^h$ for example, the above two schemes can be rewritten as follows,
\begin{equation}
  \frac{d}{dt}\sigma^h=L_h(\sigma^h;E^n).
\end{equation}
Then the TVDRK3 scheme reads, for any $n\geq 0$,
\begin{equation}
  \begin{cases}
  \sigma^{(0)}=\sigma^n, \\
  \sigma^{(1)}=\sigma^{(0)}+\tau L_h(\sigma^{(0)};E^n), \\
  \sigma^{(2)}=\frac{3}{4}\sigma^{(0)}+\frac{1}{4}\sigma^{(1)}+\frac{1}{4}\tau L_h(\sigma^{(1)};E^n), \\
  \sigma^{(3)}=\frac{1}{3}\sigma^{(0)}+\frac{2}{3}\sigma^{(2)}+\frac{2}{3}\tau L_h(\sigma^{(2)};E^n), \\
  \sigma^{n+1}=\sigma^{(3)}.
\end{cases}
\end{equation}
In each stage, we have to solve the auxiliary equation \eqref{ldg1} in the LDG scheme and need to apply the slope limiter in both schemes.

\subsection{Numerical Methods for Poisson's Equation}

Here we apply three methods to solve Poisson's equation in model \eqref{1dmodel}: finite volume method (FVM), discontinuous Galerkin (DG) method and least-squares finite element method (LSFEM).

\subsubsection{The FVM}

In this method, the numerical solution for electric potential, $\phi_j$ is defined in the center of element $I_j$. The standard second order central difference method reads,
\begin{equation}
  -\frac{\phi_{j-1}^n-2\phi_j^n+\phi_{j+1}^n}{h^2}=\rho_j^n-\sigma_j^n,\text{ for }j=0,1,\cdots,N-1,
\end{equation}
where $\rho_j^n$ and $\sigma_j^n$ are the approximate values of $\rho$ and $\sigma$ in element centers; namely, if the numerical solution of $\rho$ and $\sigma$ in the $j$-th element can be presented by $\sum_{l=0}^{p}\rho_{j,l}^nP_l(\xi)$ and $\sigma_j=\sum_{l=0}^{p}\sigma_{j,l}^nP_l(\xi)$, then $\rho_j^n=\sum_{l=0}^{p}\rho_{j,l}^nP_l(0)$ and $\sigma_j^n=\sum_{l=0}^{p}\sigma_{j,l}^nP_l(0)$. The boundary conditions are strongly imposed by introducing ghost cells and linear interpolation,
\begin{equation}\label{fdinterpolation}
  \phi_{-1}=2\phi(0,t^n)-\phi_0,\;\phi_N=2\phi(1,t^n)-\phi_{N-1}.
\end{equation}
After obtaining the numerical electric potential $\phi$, the numerical electric field at each node is defined by
\begin{equation}
  E^n|_{z_j}=\frac{\phi_{j-1}^n-\phi_j^n}{h},\text{ for }j=0,1,\cdots,N.
\end{equation}

\subsubsection{The DG method}

Define the bilinear form $B_{\epsilon} : V_k\times V_k\rightarrow\mathbb{R}$,
\begin{eqnarray}\label{dgbilinear}
B_{\epsilon}(u,v)&=&\sum_{j=0}^{N-1}\int_{I_j}\frac{du}{dz}\frac{dv}{dz}-\sum_{j\in N_i\cup N_d}\left\{\frac{du}{dz}(z_j)\right\}[v(z_j)]\nonumber \\
&&+\epsilon\sum_{j\in N_i\cup N_d}\left\{\frac{dv}{dz}(z_j)\right\}[u(z_j)]+\sum_{j\in N_i\cup N_d}\frac{\alpha}{h^{\beta}}[u(z_j)][v(z_j)],
\end{eqnarray}
and linear functional $L : V_k\rightarrow\mathbb{R}$,
\begin{equation}
  L(v)=\int_0^1(\rho^n-\sigma^n)v +\left(\epsilon\frac{dv}{dz}(z_N)+\frac{\alpha}{h^{\beta}}v(z_N)\right)\phi(1,t^n).
\end{equation}
Then the DG method is to find $\phi^h\in V_k$ such that
\begin{equation}
  B_{\epsilon}(\phi^h,v)=L(v),\;\forall v\in V_k.
\end{equation}
The DG method has different properties depending on the choice of parameters $\epsilon$, $\alpha$ and $\beta$ in \eqref{dgbilinear}. In comparison and practice, we choose $\epsilon=-1$ to form a symmetric linear system which is called symmetric interior penalty Galerkin method (SIPG) and then choose $\alpha=2$ and $\beta=1$ to ensure optimal convergence.

Comparing with the continuous Galerkin method, we introduce extra interior terms in the scheme; therefore, the electric field is approximated by
\begin{equation}
  E^n(z_j)=-\left\{\frac{d\phi^h}{dz}(z_j)\right\}+\frac{\alpha}{h^{\beta}}[\phi^h(z_j)],\text{ for }j=1,2,\cdots,N-1,
\end{equation}
and at the boundary
\begin{equation}
  \begin{aligned}
    &E^n(z_0)=-\frac{d\phi^h}{dz}(z_0)+\frac{\alpha}{h^{\beta}}(\phi(0,t^n)-\phi^h(z_0)), \\
    &E^n(z_N)=-\frac{d\phi^h}{dz}(z_N)+\frac{\alpha}{h^{\beta}}(\phi^h(z_N)-\phi(1,t^n)).
  \end{aligned}
\end{equation}

\subsubsection{The LSFEM}

This method separates the Poisson's equation into a first order differential equation system,
\begin{equation}
  \begin{cases}
  \frac{dE}{dz}=\rho^n-\sigma^n, \\
  E=-\frac{d\phi}{dz},
\end{cases}
\end{equation}
then we can treat $\phi$ and $E$ as independent variables. Usually, we call $\phi$ the scalar variable and $E$ the flux variable.

Denote the $C^0$ nodal finite element space by
$$W_k=\{v : v|_{I_j}\in \mathbb{P}_k(I_j),\text{ for }j=0,1,\cdots,N-1,\text{ v in continuous in [0,1]}\}.$$
Let $W_k^0=W_k\cap\{v : v(0)=v(1)=0\}$, $W_k^S=W_k\cap\{v : v(0)=\phi(0,t^n),v(1)=\phi(1,t^n)\}$ and $W_k^F=W_k$ be the spaces for test functions, scalar variable and flux variable respectively.

Define the bilinear form $B[(E,\phi),(w,\psi)] : (W_k^F\times W_k^S)\times(W_k^F\times W_k^0)\rightarrow\mathbb{R}$,
\begin{equation}
  B[(E,\phi),(w,\psi)]=\int_0^1\frac{dE}{dz}\frac{dw}{dz}+\int_0^1\left(E+\frac{d\phi}{dz}\right)\left(w+\frac{d\psi}{dz}\right),
\end{equation}
and linear functional $L : W_k^F\rightarrow\mathbb{R}$
\begin{equation}
  L(w)=\int_0^1(\rho^n-\sigma^n)\frac{dw}{dz}.
\end{equation}
Then the weak formulation for least-squares finite element method is to find $(E,\phi)\in(W_k^F\times W_k^S)$ such that
\begin{equation}
  B[(E,\phi),(w,\psi)]=L(w),\;\forall (w,\psi)\in(W_k^F\times W_k^0).
\end{equation}

Since the flux variable $E$ is continuous in this method, the approximate electric field is naturally chosen as $E^n=E$.

\subsection{Numerical Comparisons and Application on Double-headed Streamer Propagation}

To make our methods comparable, we choose linear polynomial approximation in DG and LSFEM method such that they are expected to have second order of accuracy. Assume the number of element in each method is $N$, then the number of unknowns in one single time step are given Table \ref{nounknown}.
\begin{table}[h]
\begin{center}
\begin{tabular}{|c|c|c|c|c|}
\hline
\multicolumn{2}{|c|}{Continuity equations} & \multicolumn{3}{|c|}{Poisson's equation} \\
\hline
OBBDG & LDG & FVM & SIPG & LSFEM \\
\hline
$3\cdot2\cdot 2N$ & $3\cdot3\cdot 2N$ & $N$ & $2N$ & $2N+2$ \\
\hline
\end{tabular}
\caption{Number of unknowns in one single time step for different methods. Note that there are three stages in TVDRK3 method and on each stage we have to solve two (three) equations for the OBBDG method (LDG method) respectively. Additionally, the Poisson's equation is assumed to be solved once in one single time step.}\label{nounknown}
\end{center}
\end{table}

Suppose the combinatorial algorithm is denoted by A+B where Method A and B are applied to solve Poisson's equation and continuity equation respectively. In the view of efficiency, we only consider four combinations: FVM+LDG, FVM+OBBDG, SIPG+OBBDG and LSFEM+OBBDG.

The comparisons are carried on a double-headed streamer propagation \cite{wu}. The gap length is 1 cm and the applied voltage is 52 kV. The gas between electrodes is nitrogen at 300K under standard atmosphere $P=760$ torr. After dimensionless, the coefficients in model \eqref{1dmodel} are $\mu_{\rho}=0.009$, $D=9.0716\times 10^{-5}$, $S=4332$, $K=-3.9315$, and initial data is set to be
  $$\sigma(z,0)=\rho(z,0)=0.0035+3.4752\times10^3\times\exp\{-[(z-0.5)/0.027]^2\}.$$

The terminal time is set to be $T=0.1$ which corresponds to 5 ns. To compare the convergence rate in space for each coupled method, the time step is chosen to be sufficient small. The results from Table \ref{Ex1test} indicate that all the four methods can be used to simulate the streamer propagation if the mesh size is smaller than some threshold.

\begin{table}[!h]
      \begin{center}
        \begin{tabular}{|c|c|c|c|c|c|c|}
        \hline
        \multicolumn{7}{|c|}{Error and convergence rate for $\sigma$.} \\
        \hline
         & & $h_0=\frac{1}{64}$ & $h_0/2$ & $h_0/2^2$ & $h_0/2^3$ & $h_0/2^4$ \\
         \hline
         FVM & error & 0.0135 & 0.0123 & 0.0035 & 4.7084E-4 & 4.1914E-5 \\
        +LDG & rate & - & 0.1322 & 1.7976 & 2.9115 & 3.4898 \\
        \hline
        FVM & error & 0.0538 & 0.0293 & 0.0086 & 0.0022 & 4.5700E-4 \\
        +OBBDG & rate & - & 0.8787 & 1.7688 & 1.9855 & 2.2463 \\
        \hline
        SIPG & error & 0.0851 & 0.0457 & 0.0134 & 0.0034 & 7.0012E-4 \\
        +OBBDG & rate & - & 0.8986 & 1.7708 & 1.9825 & 2.2739 \\
        \hline
        LSFEM & error & 0.0626 & 0.0347 & 0.0098 & 0.0024 & 4.9211E-4 \\
        +OBBDG & rate & - & 0.8497 & 1.8301 & 2.0254 & 2.2847 \\
        \hline
        \multicolumn{7}{|c|}{Error and convergence rate for $\rho$.} \\
        \hline
         & & $h_0=\frac{1}{64}$ & $h_0/2$ & $h_0/2^2$ & $h_0/2^3$ & $h_0/2^4$ \\
         \hline
         FVM & error & 0.0137 & 0.0123 & 0.0035 & 4.7227E-4 & 4.5860E-5 \\
        +LDG & rate & - & 0.1571 & 1.7973 & 2.9076 & 3.3643 \\
        \hline
        FVM & error & 0.0540 & 0.0293 & 0.0086 & 0.0022 & 4.5739E-4 \\
        +OBBDG & rate & - & 0.8827 & 1.7688 & 1.9853 & 2.2453 \\
        \hline
        SIPG & error & 0.0851 & 0.0457 & 0.0134 & 0.0034 & 7.0030E-4 \\
        +OBBDG & rate & - & 0.8987 & 1.7708 & 1.9825 & 2.2736 \\
        \hline
        LSFEM & error & 0.0627 & 0.0347 & 0.0098 & 0.0024 & 4.9212E-4 \\
        +OBBDG & rate & - & 0.8521 & 1.8302 & 2.0254 & 2.2847 \\
        \hline
        \multicolumn{7}{|c|}{Error and convergence rate for $\phi$.} \\
        \hline
         & & $h_0=\frac{1}{64}$ & $h_0/2$ & $h_0/2^2$ & $h_0/2^3$ & $h_0/2^4$ \\
         \hline
         FVM & error & 8.4380E-4 & 5.3150E-5 & 1.4045E-5 & 3.3630E-6 & 6.0467E-7 \\
        +LDG & rate & - & 3.9887 & 1.9200 & 2.0622 & 2.4756 \\
        \hline
        FVM & error & 8.2013E-4 & 5.8894E-5 & 9.3893E-6 & 1.9641E-6 & 3.7919E-7 \\
        +OBBDG & rate & - & 3.7996 & 2.6490 & 2.2571 & 2.3729 \\
        \hline
        SIPG & error & 2.2345E-4 & 3.4253E-5 & 4.9606E-6 & 1.9926E-6 & 5.1926E-7 \\
        +OBBDG & rate & - & 2.7057 & 2.7876 & 1.3159 & 1.9401 \\
        \hline
        LSFEM & error & 1.8601E-4 & 1.3564E-5 & 3.0023E-6 & 7.1625E-7 & 1.4327E-7 \\
        +OBBDG & rate & - & 3.7775 & 2.1757 & 2.0676 & 2.3217 \\
        \hline
        \multicolumn{7}{|c|}{Error and convergence rate for $E$.} \\
        \hline
         & & $h_0=\frac{1}{64}$ & $h_0/2$ & $h_0/2^2$ & $h_0/2^3$ & $h_0/2^4$ \\
         \hline
        SIPG & error & 0.0029 & 8.1787E-4 & 1.5831E-4 & 8.8592E-5 & 3.4534E-5 \\
        +OBBDG & rate & - & 1.7121 & 2.4613 & 0.8375 & 1.3592 \\
        \hline
        LSFEM & error & 0.0116 & 0.0010 & 1.0819E-4 & 1.8186E-5 & 3.2629E-6 \\
        +OBBDG & rate & - & 3.5392 & 3.2113 & 2.5727 & 2.4786 \\
        \hline
        \end{tabular}
        \caption{Numerical tests and comparisons for four different methods. It is shown that each method is acceptable since the physical quantities can obtain the theoretical convergence rate.}\label{Ex1test}
      \end{center}
    \end{table}

\begin{figure}[!h]
    \includegraphics[width=8cm]{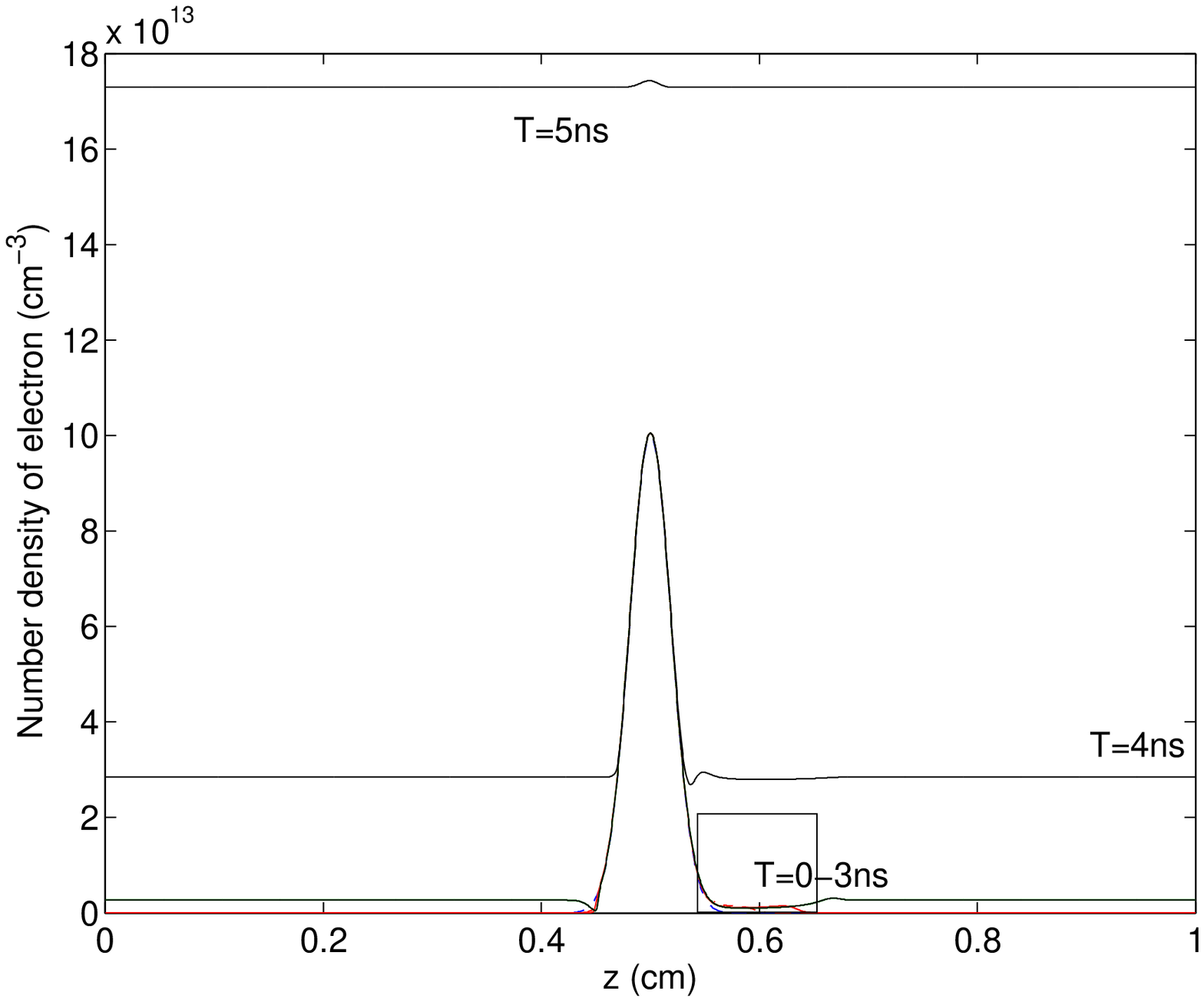}
 \includegraphics[width=8cm]{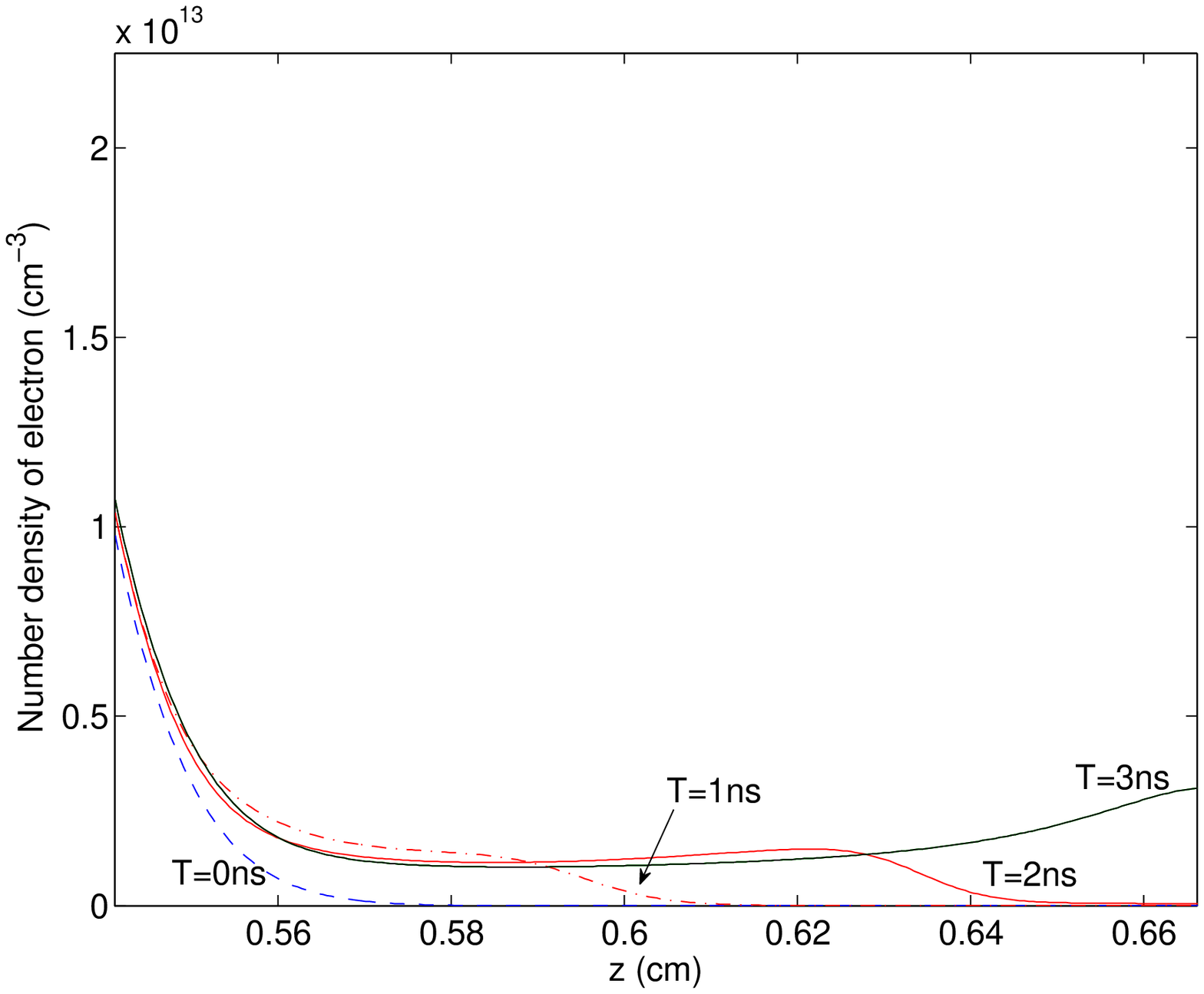}\\
  \includegraphics[width=8cm]{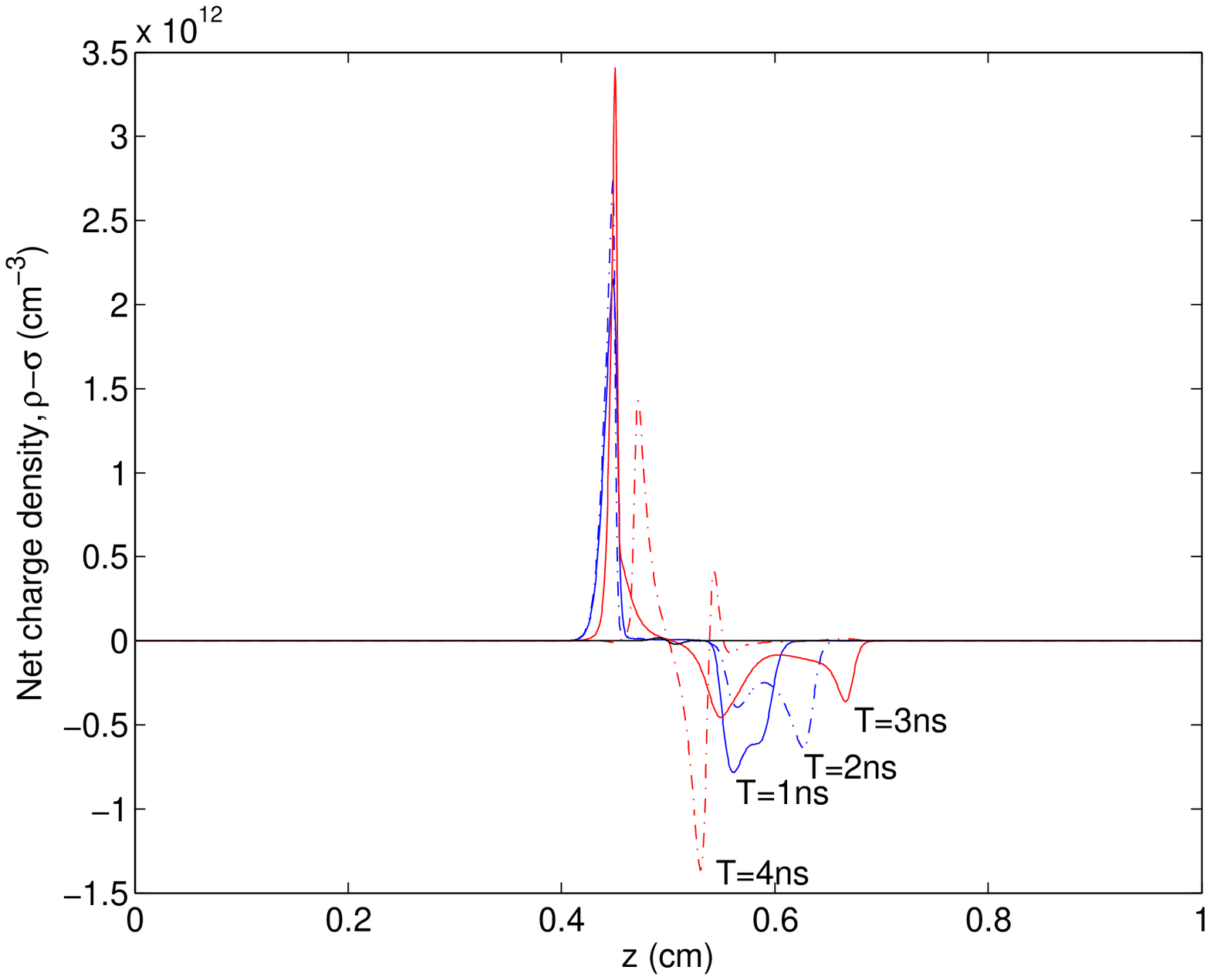}
 \includegraphics[width=8cm]{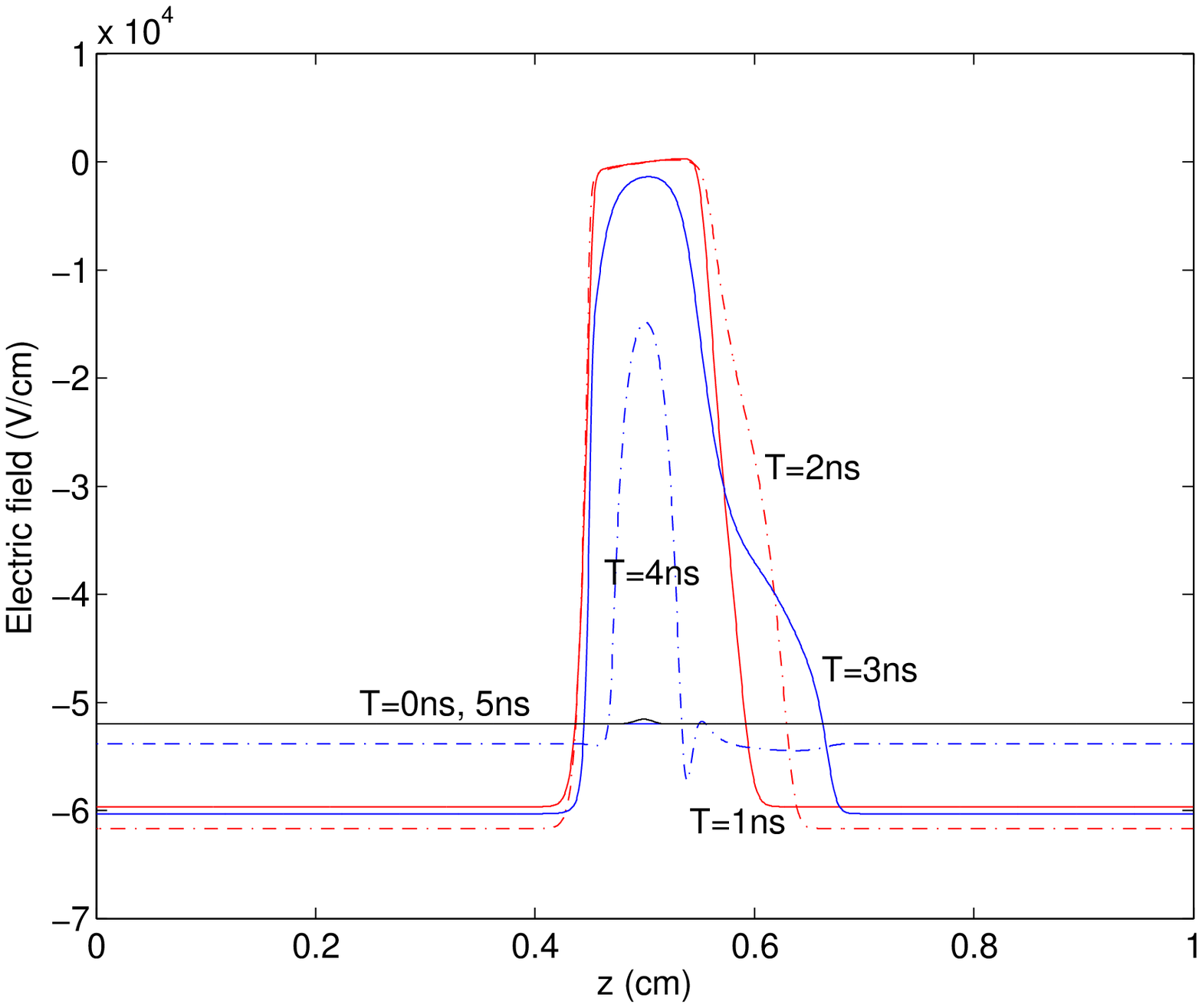}
 \caption{The profiles of a double headed streamer propagation in Example 1 at different time. The left top figure shows the number density of electron, in which the rectangular box is zoomed in and shown in the right top figure. The left bottom figure shows the number density of net charge, from which we can observe the net charge density is significantly less than the density of either electron or positive ion. The right bottom figure shows the electric field.}\label{Ex1-f1}
\end{figure}

From the previous comparison, it can be found that all of the four numerical methods are competitive candidates for solving the streamer propagation models in the view of accuracy. If the discharge region has a simple geometry, e.g. the gap between two parallel plates, FVM+LDG or FVM+OBBDG will be applied because of its easy implementation. On the other hand, if the geometry is complex, e.g. the point-to-plate gap, it is better to choose SIPG+OBBDG or LSFEM+OBBDG. In this section, since we are dealing with 1D model, we choose FVM+OBBDG to simulate the streamer propagation in 1 cm gap of nitrogen and the dynamics of the propagation process is shown in Figure \ref{Ex1-f1}.

\section{Numerical Methods in Quasi 2D model}

In this section, we focus on the behavior of the solution in model \eqref{q2dmodel}. The same numerical methods as in 1D model will be applied in this model.

Let $r^0=r_0<r_1<\cdots<r_N=1$ be a uniform spatial partition of computational domain $[r^0,1]$ such that $r_j=r^0+jh$ where $h=\frac{1-r^0}{N}$ for $j=0,1,\cdots,N$. Denote the subintervals by $I_j=[r_j,r_{j+1}]$, $j=0,1,\cdots,N-1$. The definitions of other notations are inherited from previous section.

The main difficulty in quasi 2D model is the factor $\frac{1}{r}$. When $r^0$ is closed to 0, this factor becomes singular. Suppose there is a function $u(x,y)\equiv u(\sqrt{x^2+y^2})$ defined on a 2D domain $\Omega$ with central symmtry and the test function is denoted by $v$. By applying polar coordinate to change the variable, i.e., $r=\sqrt{x^2+y^2}$, we can obtain
$$\int_{\Omega}(\triangle u)v dxdy=2\pi\int_r^R\frac{1}{r}\frac{d}{dr}\left(r\frac{du}{dr}\right)vrdr=2\pi\int_r^R\frac{1}{r}\frac{d}{dr}\left(r\frac{du}{dr}\right)\cdot(rv)dr.$$
Therefore, to overcome the singularity caused by $\frac{1}{r}$, test functions $v$ together with $r$, where $v$ is the test function in the previous section, are applied in Galerkin-type schemes.

\subsection{Discontinuous Galerkin Method for Continuity Equations}

\subsubsection{The OBBDG method}

The OBBDG method is to find $\sigma^h(r,t),\rho^h(r,t)\in V_k$ such that for $t=0$,
\begin{equation}
  \int_{r^0}^1(\sigma^h(r,0)-\sigma(r,0))rv=\int_{r^0}^1(\rho^h(r,0)-\rho(r,0))rv=0,\;\forall v\in V_k;
\end{equation}
and for $t=t^n>0$,
\begin{equation}
  \begin{aligned}
    &\int_{r^0}^1\partial_t\sigma^hrv+C(\sigma^h,v;E^n)+B(\sigma^h,v)=L(\sigma^h,v;E^n),\;\forall v\in V_k, \\
    &\int_{r^0}^1\partial_t\rho^hrv+C(\rho^h,v;E^n)=L(\sigma^h,v;E^n),\;\forall v\in V_k.
  \end{aligned}
\end{equation}
Similar as those in axial direction, the convection term is discretized by
\begin{eqnarray*}
C(P^h,v;E^n)&=&-\sum_{j=0}^{N-1}\int_{I_j}rP^h\mu_PE^n\frac{dv}{dr}+\sum_{j\in N_i}r_j\widehat{P^h}(r_j)\mu_PE^n(r_j)[v(r_j)]\nonumber \\
&+&\sum_{n\in N_n}r_j[P^h(r_j)\mu_PE^n(r_j)v(r_j)],
\end{eqnarray*}
where the numerical flux is defined by
$$\widehat{P^h}(r)=\begin{cases}
  P^h(r^-) & \text{ if }\mu_PE^n(r)\geq 0 \\
  P^h(r^+) & \text{ if }\mu_PE^n(r)< 0
  \end{cases},$$
for $P=\sigma$ or $\rho$. The diffusion term is discretized by
$$
  B(\sigma^h,v)=\sum_{j=0}^{N-1}\int_{I_j}rD\frac{d\sigma^h}{dz}\frac{dv}{dz}-\sum_{j\in N_i}r_j\left\{D\frac{d\sigma^h}{dr}(r_j)\right\}[v(r_j)]+\sum_{j\in N_i}r_j\left\{D\frac{dv}{dr}(r_j)\right\}[\sigma^h(r_j)],
$$
and the source term is discretized by
$$
  L(\sigma^h,v;E^n)=\int_{r^0}^1S|E^n|e^{K/|E^n|}\sigma^hrv.
$$

\subsubsection{The LDG method}

This method defines an auxiliary variable
$$q=\frac{\partial\sigma}{\partial r},$$
then is to find $\sigma^h(z,t),\rho^h(z,t),q^h\in V_k$ such that for $t=0$,
\begin{equation}
  \int_{r^0}^1(\sigma^h(r,0)-\sigma(r,0))rv=\int_{r^0}^1(\rho^h(r,0)-\rho(r,0))rv=0,\;\forall v\in V_k;
\end{equation}
and for $t=t^n>0$, for each element $I_j$,
\begin{equation}
  \int_{r_0}^1q^hrv=\sum_{j=0}^Nr_j\widehat{\sigma^h}(r_j)[v(r_j)]-\int_{r_0}^1r\sigma^h\frac{dv}{dr},\;\forall v\in V_k,
\end{equation}
\begin{eqnarray}
\int_{r_0}^1\partial_t\sigma^hrv&+&\sum_{j=0}^Nr_j(\mu_{\sigma}E^n(r_j)\widetilde{\sigma^h}(r_j)-D\widehat{q^h}(r_j))[v(r_j)]\nonumber \\
&&-\int_{r_0}^1r(\sigma^h\mu_{\sigma}E^n-Dq)\frac{dv}{dr}=\int_{r_0}^1S|E^n|e^{K/|E^n|}\sigma^h rv,\;\forall v\in V_k,
\end{eqnarray}
\begin{eqnarray}
\int_{r_0}^1\partial_t\rho^hrv&+&\sum_{j=0}^Nr_j\mu_{\rho}E^n(r_j)\widetilde{\rho^h}(r_j)[v(r_j)]\nonumber \\
&&-\int_{r_0}^1r\rho^h\mu_{\rho}E^n\frac{dv}{dr}=\int_{r_0}^1S|E^n|e^{K/|E^n|}\sigma^h rv,\;\forall v\in V_k.
\end{eqnarray}
The definition of numerical flux is totally the same as that in axial direction.

\subsubsection{The slope limiter}

It is expected that the cell average of numerical solution would not be changed by the slope limiter \cite{co2}. Due to the orthogonality of Legendre polynomials, the cell average is automatically preserved in 1D model. However, the cell average is computed by a weight $r$ in integration. Thus, compared with the slope limiter in 1D model, there is one more step in the slope limiter for quasi 2D model: to limit the lowest order coefficient to preserve the cell average.

\subsubsection{Fully discretized formulation}

The full discretization is still carried out by TVDRK3 method which is already illustrated in 1D model.

\subsection{Numerical Methods for Poisson's Equation}

\subsubsection{The FVM}

In this method, the numerical solution for electric potential, $\phi_j$ is defined in the center of element $I_j$. The standard second order method reads,
\begin{equation}
  -\frac{\phi_{j-1}^n-2\phi_j^n+\phi_{j+1}^n}{h^2}-\frac{\phi_{j+1}^n-\phi_{j-1}^n}{2h(r_j+\frac{h}{2})}=\rho_j^n-\sigma_j^n,\text{ for }j=0,1,\cdots,N-1,
\end{equation}
where $\rho_j^n$ and $\sigma_j^n$ are the approximate values of $\rho$ and $\sigma$ in element centers. The boundary conditions are strongly imposed by introducing ghost cells. If the boundary condition is imposed by Dirichlet type, then a linear interpolation will be used. If the boundary condition is given by Neumann type, then we use reflection. After obtaining the numerical electric potential $\phi$, the numerical electric field at each node is defined by
\begin{equation}
  E^n|_{r_j}=\frac{\phi_{j-1}^n-\phi_j^n}{h},\text{ for }j=0,1,\cdots,N.
\end{equation}

\subsubsection{The DG method}

Define the bilinear form $B_{\epsilon} : V_k\times V_k\rightarrow\mathbb{R}$,
\begin{eqnarray}
B_{\epsilon}(u,v)&=&\sum_{j=0}^{N-1}\int_{I_j}r\frac{du}{dr}\frac{dv}{dr}-\sum_{j\in N_i\cup N_d}r_j\left\{\frac{du}{dr}(r_j)\right\}[v(r_j)]\nonumber \\
&+&\epsilon\sum_{j\in N_i\cup N_d}r_j\left\{\frac{dv}{dr}(r_j)\right\}[u(r_j)]+\sum_{j\in N_i\cup N_d}r_j\frac{\alpha}{h^{\beta}}[u(r_j)][v(r_j)],
\end{eqnarray}
and linear form $L : V_k\rightarrow\mathbb{R}$,
\begin{equation}
  L(v)=\int_{r^0}^1(\rho-\sigma)r v +\sum_{j\in N_d}r_j\left(\epsilon\left[\frac{dv}{dr}(r_j)\right]+\frac{\alpha}{h^{\beta}}v(r_j)\right)\phi(r_j,t^n).
\end{equation}
Then the DG method is to find $\phi^h\in V_k$ such that
\begin{equation}
  B_{\epsilon}(\phi^h,v)=L(v),\;\forall v\in V_k.
\end{equation}
We also choose SIPG method and choose $\alpha=2$ and $\beta=1$ to ensure optimal convergence.

The electric field is approximated by
\begin{equation}
  E^n(r_j)=-\left\{\frac{d\phi^h}{dr}(r_j)\right\}+\frac{\alpha}{h^{\beta}}[\phi^h(r_j)],\text{ for }j=1,2,\cdots,N-1.
\end{equation}
At the boundary, if it is Case 1, we set
\begin{equation}
  \begin{aligned}
    &E^n(r_0)=-\frac{d\phi^h}{dr}(r_0), \\
    &E^n(r_N)=-\frac{d\phi^h}{dr}(r_N)+\frac{\alpha}{h^{\beta}}(\phi^h(r_N)-\phi(1,t^n));
  \end{aligned}
\end{equation}
if it is Case 2, then we set
\begin{equation}
  \begin{aligned}
    &E^n(r_0)=-\frac{d\phi^h}{dr}(r_0)+\frac{\alpha}{h^{\beta}}(\phi(r^0,t^n)-\phi^h(r_0)), \\
    &E^n(r_N)=-\frac{d\phi^h}{dr}(r_N)+\frac{\alpha}{h^{\beta}}(\phi^h(r_N)-\phi(1,t^n)).
  \end{aligned}
\end{equation}

\subsubsection{The mixed finite element method (MFEM)}

This method separates the Poisson's equation into a first order differential equation system \cite{br},
\begin{equation}
  \begin{cases}
  \frac{1}{r}\frac{d(rE)}{dr}=\rho^n-\sigma^n, \\
  E+\frac{d\phi}{dr}=0,
\end{cases}
\end{equation}
then we can treat $\phi$ and $E$ as independent variables. Usually, we call $\phi$ the scalar variable and $E$ the flux variable.

The reason why we use MFE rather than LSFEM comes from two simple numerical tests. In these two tests, we compare the results from continuous Galerkin method (CG), MFEM and LSFEM with linear polynomial approximation.
Example 1. Consider
$$\begin{cases}
  \frac{1}{r}\frac{d}{dr}\left(r\frac{d\phi}{dr}\right)=1, \text{ in }(0,1)\\
  \phi'(0)=0,\; \phi(1)=1,
\end{cases}$$
whose exact solution is
$$\phi=\frac{3}{4}+\frac{1}{4}r^2.$$
The comparison are shown in Table \ref{mfe12}.

\begin{table}[!h]
      \begin{center}
        \begin{tabular}{|c|c|c|c|c|c|}
        \hline
        \multicolumn{6}{|c|}{Error and convergence order for $\phi$.} \\
        \hline
         & & $h_0=1/16$ & $h_0/2$ & $h_0/2^2$ & $h_0/2^3$ \\
         \hline
        CG & error & 9.6305E-5 & 2.4079E-5 & 6.0190E-6 & 1.5046E-6 \\
         & order & - & 1.9998 & 2.0002 & 2.0001 \\
         \hline
        MFE & error & 0.0045 & 0.0023 & 0.0011 & 5.6381E-4 \\
         & order & - & 0.9993 & 0.9998 & 1.0000 \\
         \hline
        LS & error & 9.8487E-5 & 2.4628E-5 & 6.1566E-6 & 1.5390E-6 \\
         & order & - & 1.9996 & 2.0001 & 2.0001 \\
         \hline
         \multicolumn{6}{|c|}{Error and convergence order for $E=-\phi'$.} \\
        \hline
         & & $h_0=1/16$ & $h_0/2$ & $h_0/2^2$ & $h_0/2^3$ \\
         \hline
        CG & error & 0.0064 & 0.0032 & 0.0016 & 7.9733E-4 \\
         & order & - & 0.9984 & 0.9995 & 0.9999 \\
         \hline
        MFE & error & 7.0705E-18 & 7.6120E-18 & 1.4591E-17 & 1.0693E-17 \\
         & order & - & - & - & - \\
         \hline
         LS & error & 1.4365E-5 & 3.5952E-6 & 8.9905E-7 & 2.2478E-7 \\
         & order & - & 1.9984 & 1.9996 & 1.9999 \\
         \hline
         \end{tabular}
         \caption{The comparisons for Example 1.}\label{mfe12}
         \end{center}
         \end{table}

Example 2. Consider
$$\begin{cases}
  \frac{1}{r}\frac{d}{dr}\left(r\frac{d\phi}{dr}\right)=0, \text{ in }(0.05,1)\\
  \phi(0.05)=0,\; \phi(1)=1,
\end{cases}$$
whose exact solution is
$$\phi=1-\frac{\ln r}{\ln0.05}.$$
The comparison are shown in Table \ref{mfe22}.

\begin{table}[!h]
      \begin{center}
        \begin{tabular}{|c|c|c|c|c|c|}
        \hline
        \multicolumn{6}{|c|}{Error and convergence order for $\phi$.} \\
        \hline
         & & $h_0=1/16$ & $h_0/2$ & $h_0/2^2$ & $h_0/2^3$ \\
         \hline
        CG & error & 0.0032 & 8.8972E-4 & 2.3011E-4 & 5.8079E-5 \\
         & order & - & 1.8555 & 1.9510 & 1.9862 \\
         \hline
        MFE & error & 0.0143 & 0.0057 & 0.0026 & 0.0013 \\
         & order & - & 1.3264 & 1.1446 & 1.0442 \\
         \hline
        LS & error & 0.0132 & 0.0047 & 0.0013 & 3.4884E-4 \\
         & order & - & 1.4942 & 1.8046 & 1.9426 \\
         \hline
         \multicolumn{6}{|c|}{Error and convergence order for $E=-\phi'$.} \\
        \hline
         & & $h_0=1/16$ & $h_0/2$ & $h_0/2^2$ & $h_0/2^3$ \\
         \hline
        CG & error & 0.0746 & 0.0394 & 0.0201 & 0.0101 \\
         & order & - & 0.9216 & 0.9737 & 0.9926 \\
         \hline
        MFE & error & 0.0158 & 0.0043 & 0.0011 & 2.7835E-4 \\
         & order & - & 1.8729 & 1.9624 & 1.9900 \\
         \hline
         LS & error & 0.0542 & 0.0195 & 0.0056 & 0.0014 \\
         & order & - & 1.4757 & 1.8122 & 1.9475 \\
         \hline
         \end{tabular}
         \caption{The comparisons for Example 2.}\label{mfe22}
         \end{center}
         \end{table}

Table \ref{mfe12} and \ref{mfe22} suggest that MFEM is the best method for both of Case 1 and Case 2 if we expect that the numerical solution of $E=-\frac{d\phi}{dr}$ is as more accuracy as possible. LSFEM has the same order of accuracy for $E$ as MFE, however, it requires a more finer mesh.

Denote the $C^0$ nodal finite element space by
$$W_k=\{v : v|_{I_j}\in \mathbb{P}_k(I_j),\text{ for }j=0,1,\cdots,N-1. v \text{ is continuous in } [r^0,1]\}.$$
Let $W_k^1=W_k\cap\{v : v(r^0)=0\}$ or $W_k^2=W_k$ be the space for flux variable in Case 1 or Case 2. Due to stability, the space for scalar variable should be determined by inf-sup condition. For example if we choose $k=1$, the space for scalar variable is identically equal to $V_0$ in previous section.

The weak formulation for MFEM is to find $(E,\phi)\in W_1^1\times V_0$ such that
\begin{equation}
  \begin{cases}
-\int_{r^0}^1\psi\frac{d(rE)}{dr}=-\int_{r^0}^1(\rho-\sigma)r\psi,\;\forall \psi\in V_0, \\
  \int_{r^0}^1rEw+\phi(1,t^n)w(1)-\int_{r^0}^1\phi\frac{d(rw)}{dr}=0,\;\forall w\in W_1^1.
\end{cases}
\end{equation}
or to find $(E,\phi)\in W_1^2\times V_0$ such that
\begin{equation}
  \begin{cases}
-\int_{r^0}^1\psi\frac{d(rE)}{dr}=-\int_{r^0}^1(\rho^n-\sigma^n)r\psi,\;\forall \psi\in V_0, \\
  \int_{r^0}^1rEw+\phi(1,t^n)w(1)-r^0\phi(r^0,t^n)w(r^0)-\int_{r^0}^1\phi\frac{d(rw)}{dr}=0,\;\forall w\in W_1^2.
\end{cases}
\end{equation}

Since the flux variable $E$ is continuous in this method, the approximate electric field is naturally chosen as $E^n=E$.

\subsection{Numerical Comparisons and Applications}

In the view of efficiency, by regarding Table \ref{nounknown} again, we only consider four combinations: FVM+LDG, FVM+OBBDG, SIPG+OBBDG and MFEM+OBBDG as well.

\subsubsection{Accuracy test 1: $r^0=0$}

Under this configuration, the initial data for continuity equations is well separated to avoid constant initial solution to Poisson's equation; otherwise, the solutions of continuity equations would not have significant difference from initial data. There is no real experiment satisfying this requirement, only accuracy test for different methods is shown. Note that, this test is used to compare different strategies and study the extensions to quasi three dimensional model \cite{zhuang2}.
The dimensionless parameters are set by \cite{wu},
  $$\mu_{\sigma}=-2,\;\mu_{\rho}=-1,\;D=10^{-4},\;S=1000,\;K=-5;$$
  initial data is,
  $$\sigma(r,0)=\exp\{-100r^2\},\;\rho(r,0)=\exp\{-100(r-1)^2\};$$
  and the Dirichlet boundary condition at the right endpoint is
  $$\phi(1,t)=0.$$
The terminal time is $T=0.5$. The comparisons from Table \ref{Ex3test} indicate that all the four methods work. Besides, the requirement of mesh size is not so strict as that in axial direction.

\begin{table}[!h]
      \begin{center}
        \begin{tabular}{|c|c|c|c|c|c|c|}
        \hline
        \multicolumn{7}{|c|}{Error and convergence rate for $\sigma$.} \\
        \hline
         & & $h_0=\frac{1}{32}$ & $h_0/2$ & $h_0/2^2$ & $h_0/2^3$ & $h_0/2^4$ \\
         \hline
         FVM & error & 0.0010 & 2.7767E-4 & 8.2409E-5 & 2.4836E-5 & 6.5115E-6 \\
        +LDG & rate & - & 1.9098 & 1.7525 & 1.7304 & 1.9314 \\
        \hline
        FVM & error & 9.5691E-4 & 2.0306E-4 & 5.0001E-5 & 1.2706E-5 & 2.9361E-6 \\
        +OBBDG & rate & - & 2.2365 & 2.0219 & 1.9764 & 2.1136 \\
        \hline
        SIPG & error & 9.2227E-4 & 1.8626E-4 & 4.5483E-5 & 1.1488E-5 & 2.7061E-6 \\
        +OBBDG & rate & - & 2.3079 & 2.0339 & 1.9852 & 2.0859 \\
        \hline
        MFEM & error & 9.2247E-4 & 1.8617E-4 & 4.4937E-5 & 1.1317E-5 & 2.6420E-6 \\
        +OBBDG & rate & - & 2.3089 & 2.0506 & 1.9895 & 2.0987 \\
        \hline
        \multicolumn{7}{|c|}{Error and convergence rate for $\rho$.} \\
        \hline
         & & $h_0=\frac{1}{32}$ & $h_0/2$ & $h_0/2^2$ & $h_0/2^3$ & $h_0/2^4$ \\
         \hline
         FVM & error & 0.0182 & 0.0049 & 0.0014 & 3.5623E-4 & 8.0545E-5 \\
        +LDG & rate & - & 1.8878 & 1.8092 & 1.9764 & 2.1449 \\
        \hline
        FVM & error & 0.0182 & 0.0049 & 0.0014 & 3.5623E-4 & 8.0545E-5 \\
        +OBBDG & rate & - & 1.8877 & 1.8093 & 1.9764 & 2.1449 \\
        \hline
        SIPG & error & 0.0185 & 0.0049 & 0.0014 & 3.4212E-4 & 7.8791E-5 \\
        +OBBDG & rate & - & 1.9302 & 1.8151 & 2.0137 & 2.1184 \\
        \hline
        MFEM & error & 0.0174 & 0.0048 & 0.0013 & 3.3525E-4 & 7.6774E-5 \\
        +OBBDG & rate & - & 1.8663 & 1.8267 & 2.0078 & 2.1266 \\
        \hline
        \multicolumn{7}{|c|}{Error and convergence rate for $\phi$.} \\
        \hline
         & & $h_0=\frac{1}{32}$ & $h_0/2$ & $h_0/2^2$ & $h_0/2^3$ & $h_0/2^4$ \\
         \hline
         FVM & error & 1.4470E-4 & 4.5521E-5 & 1.2672E-5 & 3.3403E-6 & 8.6001E-7 \\
        +LDG & rate & - & 1.6684 & 1.8449 & 1.9236 & 1.9576 \\
        \hline
        FVM & error & 1.4469E-4 & 4.5524E-5 & 1.2672E-5 & 3.3403E-6 & 8.6001E-7 \\
        +OBBDG & rate & - & 1.6683 & 1.8449 & 1.9236 & 1.9576 \\
        \hline
        SIPG & error & 0.0130 & 0.0029 & 7.0998E-4 & 1.7816E-4 & 4.0111E-5 \\
        +OBBDG & rate & - & 2.1495 & 2.0500 & 1.9946 & 2.1511 \\
        \hline
        MFEM & error & 0.0347 & 0.0171 & 0.0085 & 0.0041 & 0.0018 \\
        +OBBDG & rate & - & 1.0223 & 1.0138 & 1.0362 & 1.1611 \\
        \hline
       \multicolumn{7}{|c|}{Error and convergence rate for $E$.} \\
        \hline
         & & $h_0=\frac{1}{32}$ & $h_0/2$ & $h_0/2^2$ & $h_0/2^3$ & $h_0/2^4$ \\
         \hline
        SIPG & error & 0.0551 & 0.0225 & 0.0092 & 0.0039 & 0.0016 \\
        +OBB & rate & - & 1.2934 & 1.2861 & 1.2309 & 1.2898 \\
        \hline
        MFEM & error & 0.0057 & 0.0015 & 3.7874E-4 & 9.1864E-5 & 1.9365E-5 \\
        +OBBDG & rate & - & 1.9133 & 1.9929 & 2.0437 & 2.2461 \\
        \hline
        \end{tabular}
        \caption{Numerical tests and comparisons for four different methods. It is shown that each method is acceptable since the physical quantities can obtain the theoretical convergence rate.}\label{Ex3test}
      \end{center}
    \end{table}

\subsubsection{Accuracy test 2: $r^0>0$}
 A streamer propagation between coaxial circles reflects the discharge between a thin conductor and a
cylinder. The radius of outer cylinder is 1 cm and the radius of inner conductor is 1 mm. A high negative voltage of $-6.6$ kV is applied to the wire to generate discharge. Thus, the boundary conditions for Poisson's equation are imposed by,
  $$\phi(0.1,t)=-1,\;\phi(1,t)=0.$$
  The other dimensionless parameters are $D_r=2190\, {\text{cm}^2}/{\text{s}}$\cite{wu}.
  The initial data is concentrated around the wire,
  $$N_e(r,0)=N_p(r,0)=10^8+10^{14}\exp\{-[(r-0.1)/0.021]^2\} \text{ cm}^{-3}.$$

  The terminal time is $T=0.1$ which corresponds to 10 ns. To compare the convergence rate in space for each coupled method, the time step is chosen small enough. The comparisons from Table \ref{Ex4test} indicate that all the four methods can be used to simulate the streamer propagation. Now the mesh size should be smaller enough to obtain the optimal rate of convergence compared with the example in section 3.3.1.

\begin{table}[!h]
      \begin{center}
        \begin{tabular}{|c|c|c|c|c|c|c|}
        \hline
        \multicolumn{7}{|c|}{Error and convergence rate for $\sigma$.} \\
        \hline
         & & $h_0=\frac{1}{64}$ & $h_0/2$ & $h_0/2^2$ & $h_0/2^3$ & $h_0/2^4$ \\
         \hline
         FVM & error & 0.0047 & 0.0011 & 2.6046E-4 & 6.0794E-5 & 1.2759E-5 \\
        +LDG & rate & - & 2.0452 & 2.1217 & 2.0991 & 2.2524 \\
        \hline
        FVM & error & 0.0047 & 0.0013 & 2.5982E-4 & 5.5842E-5 & 1.7109E-5 \\
        +OBBDG & rate & - & 1.8396 & 2.3270 & 2.2181 & 1.7066 \\
        \hline
        SIPG & error & 0.0048 & 0.0012 & 2.9612E-4 & 7.4108E-5 & 1.8088E-5 \\
        +OBBDG & rate & - & 1.9757 & 2.0412 & 1.9985 & 2.0346 \\
        \hline
        MFEM & error & 0.0047 & 0.0013 & 2.6153E-4 & 5.5835E-5 & 1.7103E-5 \\
        +OBBDG & rate & - & 1.8569 & 2.3248 & 2.2277 & 1.7069 \\
        \hline
        \multicolumn{7}{|c|}{Error and convergence rate for $\rho$.} \\
        \hline
         & & $h_0=\frac{1}{64}$ & $h_0/2$ & $h_0/2^2$ & $h_0/2^3$ & $h_0/2^4$ \\
         \hline
         FVM & error & 0.0045 & 0.0012 & 3.8231E-4 & 1.5344E-4 & 5.7398E-5 \\
        +LDG & rate & - & 1.9324 & 1.6327 & 1.3171 & 1.4186 \\
        \hline
        FVM & error & 0.0045 & 0.0012 & 3.7991E-4 & 1.5315E-4 & 5.8506E-5 \\
        +OBBDG & rate & - & 1.9388 & 1.6349 & 1.3107 & 1.3883 \\
        \hline
        SIPG & error & 0.0047 & 0.0012 & 3.0958E-4 & 8.3709E-5 & 2.1456E-5 \\
        +OBBDG & rate & - & 1.9590 & 1.9773 & 1.8869 & 1.9640 \\
        \hline
        MFEM & error & 0.0045 & 0.0010 & 2.2878E-4 & 5.2578E-5 & 1.6734E-5 \\
        +OBBDG & rate & - & 2.1088 & 2.1889 & 2.1214 & 1.6516 \\
        \hline
        \multicolumn{7}{|c|}{Error and convergence rate for $\phi$.} \\
        \hline
         & & $h_0=\frac{1}{64}$ & $h_0/2$ & $h_0/2^2$ & $h_0/2^3$ & $h_0/2^4$ \\
         \hline
         FVM & error & 0.0221 & 0.0067 & 0.0012 & 9.6402E-5 & 1.9800E-5 \\
        +LDG & rate & - & 1.7219 & 2.5164 & 3.5993 & 2.2836 \\
        \hline
        FVM & error & 0.0236 & 0.0068 & 0.0013 & 1.3337E-4 & 2.7401E-5 \\
        +OBBDG & rate & - & 1.8050 & 2.4268 & 3.2377 & 2.2831 \\
        \hline
        SIPG & error & 0.0134 & 0.0039 & 8.7572E-4 & 1.4746E-5 & 2.5383E-5 \\
        +OBBDG & rate & - & 1.7614 & 2.1721 & 2.5701 & 2.5384 \\
        \hline
        MFEM & error & 0.0100 & 0.0035 & 0.0013 & 5.4734E-4 & 2.4045E-4 \\
        +OBBDG & rate & - & 1.5293 & 1.4667 & 1.2006 & 1.1867 \\
        \hline
        \multicolumn{7}{|c|}{Error and convergence rate for $E$.} \\
        \hline
         & & $h_0=\frac{1}{64}$ & $h_0/2$ & $h_0/2^2$ & $h_0/2^3$ & $h_0/2^4$ \\
         \hline
        SIPG & error & 0.0498 & 0.0166 & 0.0042 & 0.0010 & 3.8263E-4 \\
        +OBBDG & rate & - & 1.5875 & 1.9940 & 1.9895 & 1.4536 \\
        \hline
        MFEM & error & 0.0341 & 0.0112 & 0.0027 & 5.2027E-4 & 1.0154E-4 \\
        +OBBDG & rate & - & 1.6056 & 2.0431 & 2.3865 & 2.3573 \\
        \hline
        \end{tabular}
        \caption{Numerical tests and comparisons for four different methods. It is shown that each method is acceptable since the physical quantities can obtain the theoretical convergence rate.}\label{Ex4test}
      \end{center}
    \end{table}

\begin{figure}[!hbt]
    \includegraphics[width=8cm]{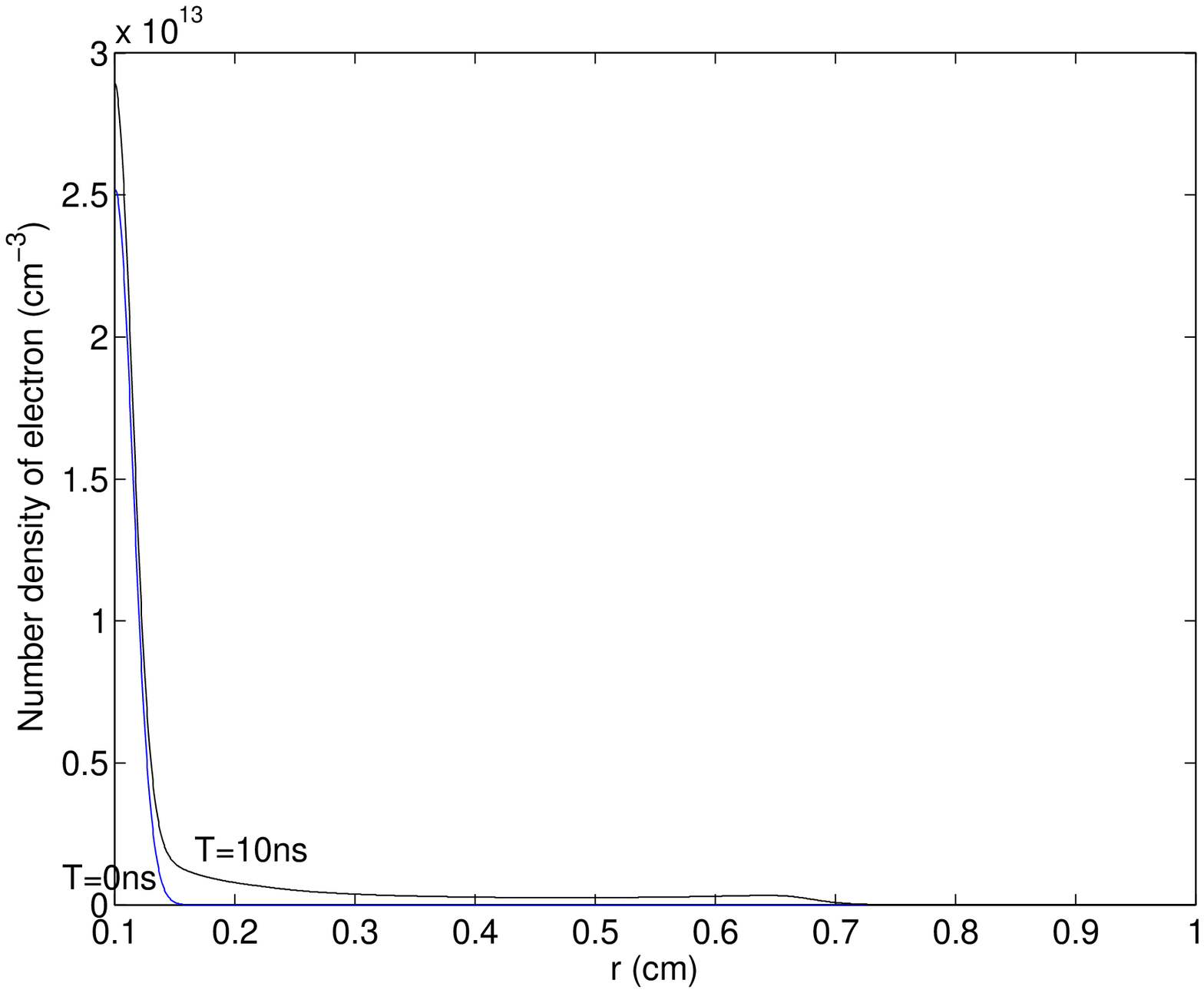}
 \includegraphics[width=8cm]{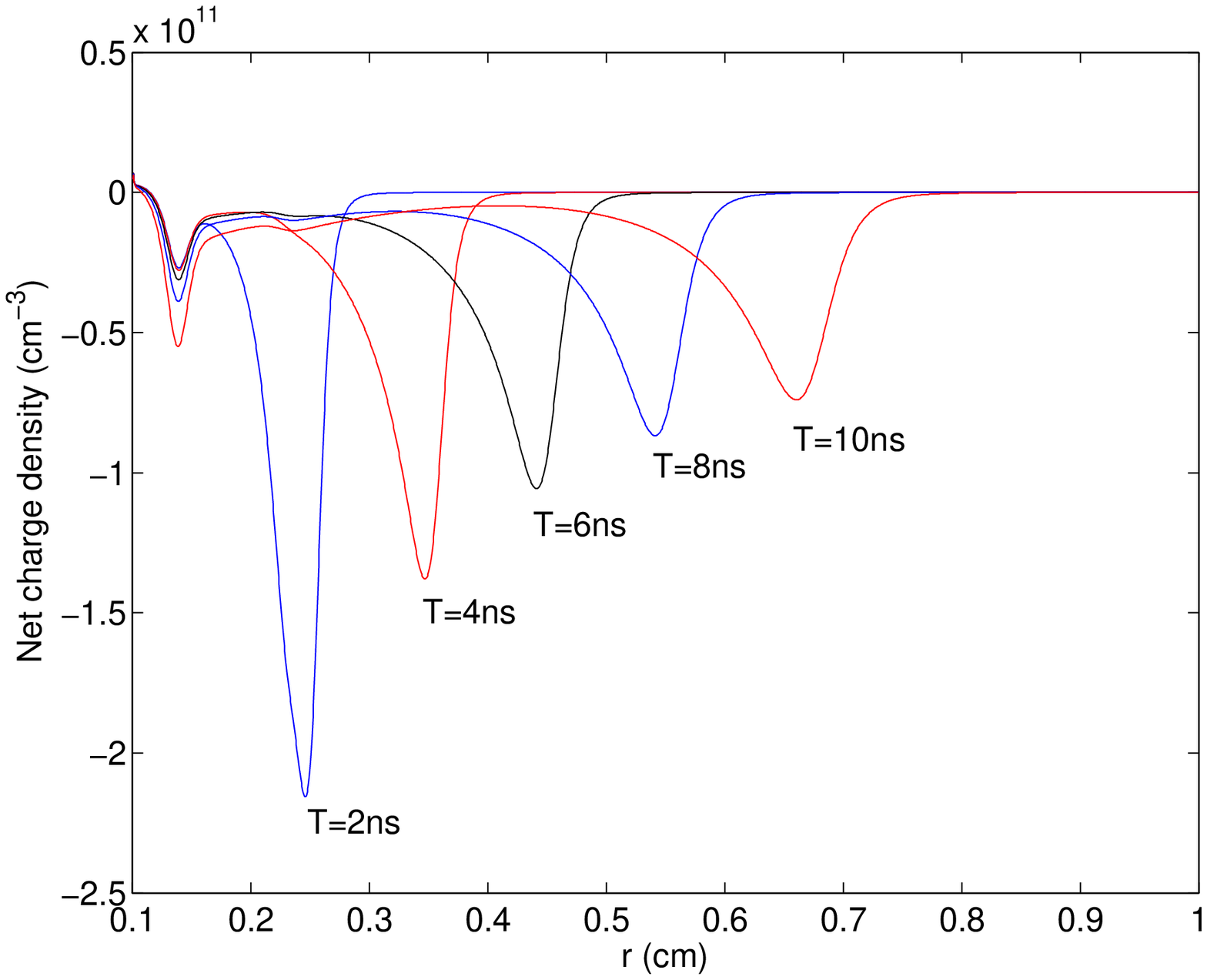}\\
  \includegraphics[width=8cm]{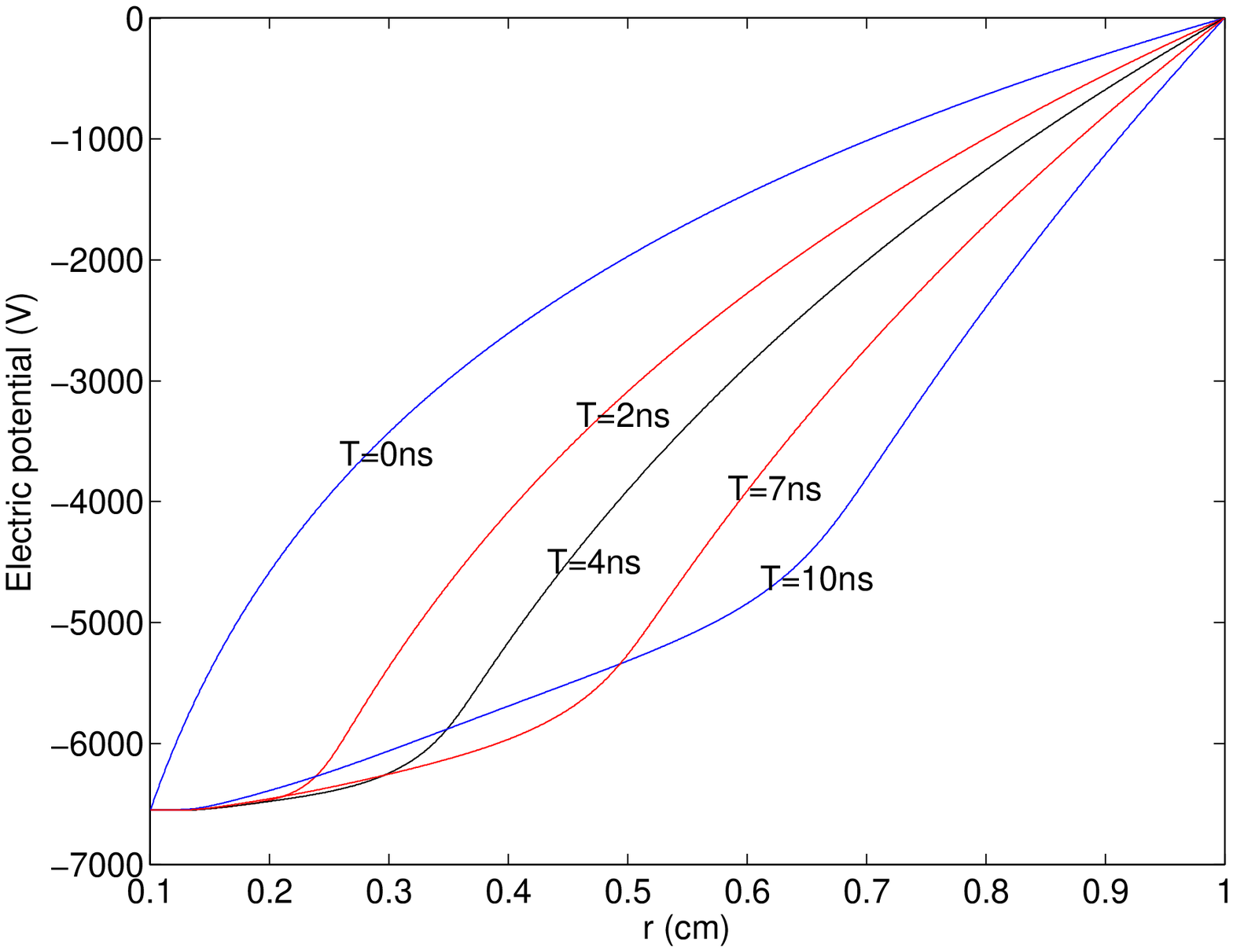}
 \includegraphics[width=8cm]{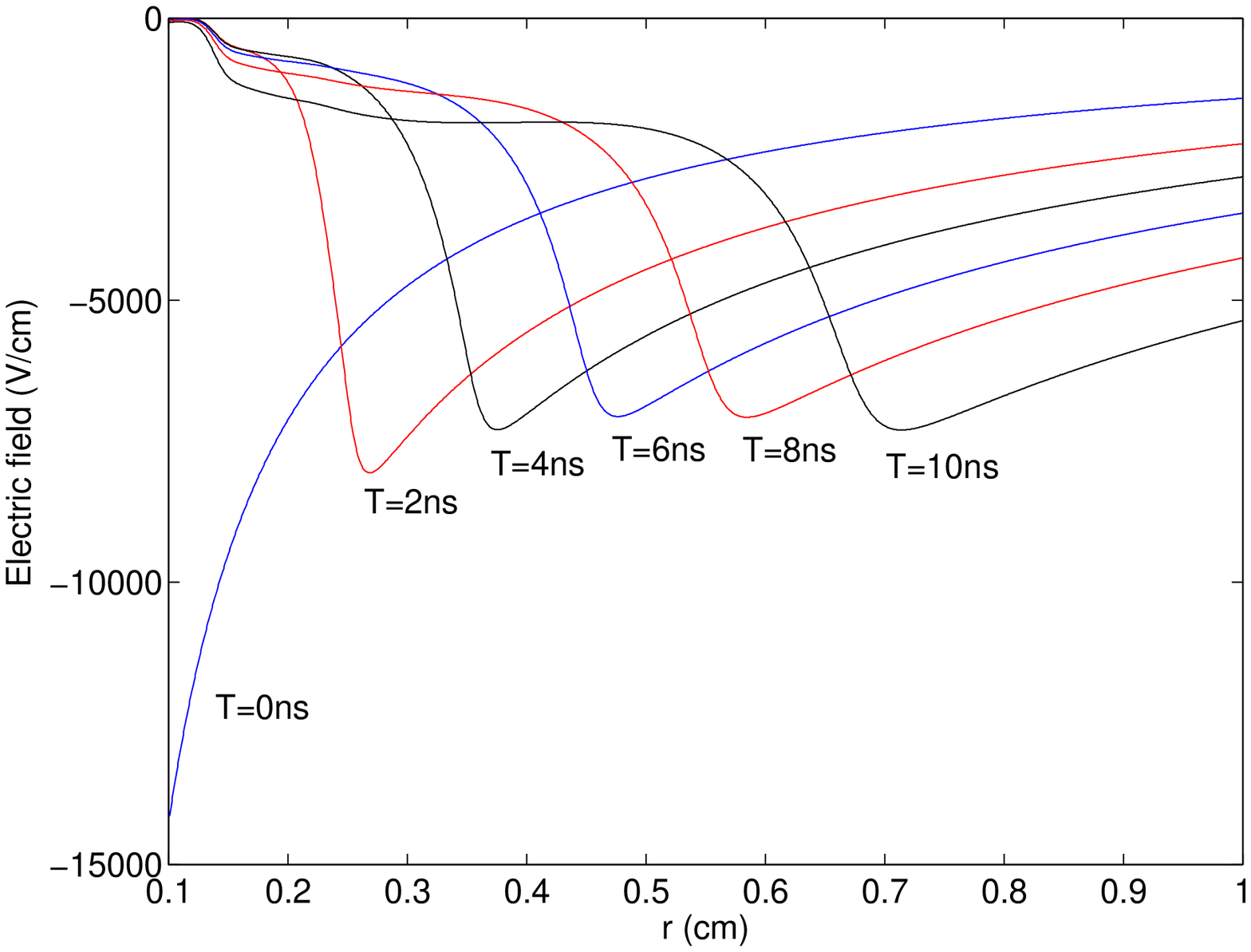}
 \caption{The profiles of a double headed streamer propagation in Example 1 at different time. The left top figure shows the number density of electron and the right top figure shows the number density of net charge, from which we can observe the net charge density is significantly less than the density of either electron or positive ion. The bottom figures show the electric potential and field.}\label{Ex4-f1}
\end{figure}

The dynamics of streamer propagation between coaxial circles are shown in Figure \ref{Ex4-f1}.

\section{Conclusions}

In this paper, we have proposed and tested four different combinatorial methods for solving the 1D and quasi 2D streamer propagation models. It is the first time that discontinuous Galerkin methods and least-squares finite element method are used to solve the Poisson's equation in the fluid model of streamer discharge.

From the numerical comparisons, it is concluded that the four combinations are compatible in the sense that all the rescaled physical quantities can attain the expected rate of convergence in each combinatorial methods as long as the spatial step length is sufficient small such that the propagation phenomenon can be resolved. It is shown through numerical examples, that DG method is indeed a competitor in simulation of streamer propagation.

The work to extend the methods to two-dimensional model and quasi three-dimensional model would be reported later.

\section*{Acknowledgement}
This work is supported by National Natural Science Foundation of China under grant 51577098 and 51207078.

\end{document}